\documentclass[12pt,a4paper]{article}
\usepackage[T2A,T1]{fontenc}
\usepackage{amssymb,amsmath,stmaryrd,mathtools,bm,graphicx,caption,subcaption,here,parskip,tikz-cd,tikz,cancel,ulem,cite,titlesec,enumerate,empheq,changepage,physics,cancel}
\usepackage[rightcaption]{sidecap}
\usepackage[margin=1in]{geometry}
\usepackage[most]{tcolorbox}
\usepackage[colorlinks,allcolors=blue]{hyperref}
\usepackage[symbol]{footmisc}

\numberwithin{equation}{section}
\DeclareMathAlphabet{\mathscrbf}{OMS}{mdugm}{b}{n}
\usetikzlibrary{decorations.markings,decorations.pathreplacing}
\newtcbox{\othermathbox}[1][]{nobeforeafter, math upper, tcbox raise base, 
          enhanced, rounded corners, colback=black!5, colframe=black,
          left=0.7em, top=0.4em, right=0.7em, bottom=0.5em}

\definecolor{MyYellow}{RGB}{248,199,82}
\usepackage[utf8]{inputenc} 
\usepackage{amsmath}

\makeatletter
\newcommand{\ostar}{\mathbin{\mathpalette\make@circled\star}}
\newcommand{\make@circled}[2]{%
  \ooalign{$\m@th#1\smallbigcirc{#1}$\cr\hidewidth$\m@th#1#2$\hidewidth\cr}%
}
\newcommand{\smallbigcirc}[1]{%
  \vcenter{\hbox{\scalebox{0.77778}{$\m@th#1\bigcirc$}}}%
}
\makeatother

\renewcommand{\i}[1]{\textit{#1}}

\newcommand{\be}{\begin{equation}}
\newcommand{\ee}{\end{equation}}

\newcommand{\der}{\partial}

\newcommand{\ie}{\textit{i.e.}\ }
\newcommand{\eg}{\textit{e.g.}\ }

\newcommand{\cH}{{\cal H}}

\newcommand{\cO}{{\cal O}}

\newcommand{\CC}{\mathbb{C}}

\newcommand{\RR}{\mathbb{R}}

\newcommand{\ZZ}{\mathbb{Z}}

\renewcommand\i[1]{\textit{#1}}

\begin{document}

\begin{center}
{\LARGE{{Dressed States Call for\\[.4em]
Logarithmic Asymptotic Symmetries}}}\\[1.3em]

{\large Jorrit Bosma,$^1$ Jeremy A. Mann,$^2$ Charles Marteau,$^2$\footnote[1]{Now at Google DeepMind, USA.}\\[.2em]
Blagoje Oblak,$^3$ and Marios Petropoulos$^2$}\\[1.3em]

{\small%
$^1$ Institute for Theoretical Physics, ETH Zürich, CH 8093 Z\"urich, Switzerland\\
$^2$ CPHT, CNRS, École polytechnique, Institut Polytechnique de Paris, 91120 Palaiseau, France\\
$^3$ Université Claude Bernard Lyon 1, ICJ UMR 5208, CNRS, 69622 Villeurbanne, France}
\end{center}
~\\[2em]

\begin{center}
\begin{minipage}{.8\textwidth}
\textbf{Abstract.} Inspired by Wigner's classification of elementary particles as irreducible unitary representations of a spacetime symmetry group, we ask what group can give rise in this way to the quantum states of a particle dressed with clouds of infrared gauge bosons. We show that the answer is given by standard asymptotic symmetries (such as BMS in the gravitational case), supplemented by the logarithmic symmetries identified in \cite{Fuentealba:2023syb}. A corollary is the factorization of the Hilbert space of a dressed particle as a tensor product of the space of `naked' one-particle states with a Hilbert space of soft gauge bosons. The latter cannot be obtained without logarithmic transformations, which are ultimately responsible for the presence of a crucial Heisenberg central extension.
\end{minipage}
\vspace{3em}

Pre-print number: CPHT-RR019.07202
\end{center}

\newpage
\section{Introduction and outline}
\label{se1}

Gravitation and gauge field theories have a rich infrared structure: they admit intricate, often infinite-dimensional asymptotic symmetries at large distances \cite{Bondi,Sachs1,Sachs2,Brown,Barnich:2009se,Barnich:2010eb,Barnich:2011mi,He:2014cra,Strominger:2017zoo}. These global symmetries are deeply related to potentially observable memory effects \cite{Zeldovich,Braginsky:1986ia,Thorne:1992sdb,Bieri:2011zb}, and the corresponding Ward identities are tantamount to soft theorems for scattering amplitudes involving low-frequency gauge bosons \cite{Strominger:2013jfa,He:2014laa,He:2014cra,Agrawal_SSB}. Relatedly, the dressed states that cure the infrared divergences of the S-matrix are eigenvectors of the conserved charges~\cite{Gabai:2016kuf,Kapec:2017tkm,Nande:2017dba,Choi:2017ylo,Gomes:2018shn,Hirai:2019gio,Hirai:2020kzx,Hirai:2022yqw,Furugori:2020vdl}. We show here that dressed states are irreducible unitary representations of the asymptotic symmetry group, provided suitable `canonically conjugate', so-called \i{logarithmic} asymptotic symmetries \cite{Fuentealba:2023syb}, are taken into account. Remarkably, such extensions of asymptotic symmetries were overlooked for decades until a series of recent publications \cite{Fuentealba:2022xsz,Fuentealba:2023rvf,Fuentealba:2025ekj,Gonzalez:2023yrz} that all converge towards the same structure.

It is worth stressing right away that the solution to the infrared problem in gauge theories is well established, irrespective of asymptotic symmetries. Following the seminal insight of Dirac \cite{Dirac:1955uv}, the key is to realize that gauge-invariant charged quantum states can only be built by `dressing' a naked charged particle with a coherent superposition of soft gauge bosons \cite{Kulish}. This ultimately yields S-matrix elements that are finite, \ie free of infrared divergences \cite{Chung:1965zza,Kibble:1968sfb,Kibble:1968oug,Kibble:1968npb,Kibble:1968lka}. From that perspective, the point of the present letter is not to cure infrared divergences, but to understand what asymptotic symmetry groups---if any---suffice to reproduce Hilbert spaces of dressed states.

Let us phrase this question in representation-theoretic language. Inspired by Wigner's work on representation of the Poincar\'e group \cite{Wigner:1939cj}, we view the irreducible unitary representations of any spacetime symmetry group as a definition of the corresponding allowed `particles'. The problem of dressed states thus becomes to find a group whose irreducible, unitary representations contain the usual relativistic particles, while also `dressing' each such naked particle by a Fock space's worth of zero-frequency gauge bosons. In this puzzle, both unitarity and irreducibility are essential, forcing the group to be at once small enough that each of its irreducible representations contains a single hard particle, yet large enough that its irreducible representations also contain clouds of soft gauge bosons. We will see that it is the criterion of irreducibility that ultimately requires the presence of extended, dual symmetries.

Just like the concept of dressed states, the idea to define particles in scattering states as representations of asymptotic symmetries is not new. It goes back to the seminal work of McCarthy on the Bondi-Metzner-Sachs (BMS) group \cite{Mccarthy:1972ry}, although we will argue that this construction is far off the mark when it comes to describing dressed states. Simply put, the induced representations built in \cite{McCarthy00,McCarthy01,McCarthy317,McCarthy301} (and recently revisited in \cite{Bekaert:2024uuy,Bekaert:2025kjb,Ruzziconi:2026isv}) are too small. What is needed instead is a group extension---central or otherwise---that makes the group structure rich enough to accommodate Fock spaces of soft gauge bosons. To some extent, an analogy can be drawn here with $(2+1)$-dimensional gravity, where central extensions of asymptotic symmetries \cite{Brown,Barnich:2006av} are indeed crucial to describe boundary gravitons and their quantization \cite{Barnich1,Barnich2,BarnichGeom,OblakThesis,Cotler}; imagine trying to tackle AdS$_3$/CFT$_2$ without knowing the Brown-Henneaux central charge! Similar extensions have in fact been put forward for $(3+1)$-dimensional symmetries \cite{Barnich:2009se,Barnich:2010eb,Barnich:2011mi}, but the difficulty of defining surface charges in the presence of radiation often makes the status of such proposals much less robust than in $(2+1)$ dimensions (where the issue of radiation does not occur).

Our arguments here are based on what is perhaps the simplest kind of central extension---even more basic than that of the Virasoro algebra in $(2+1)$ gravity. Indeed, several recent works have highlighted the appearance of additional symmetries that are canonically conjugate to standard asymptotic symmetries \cite{Fuentealba:2022xsz,Fuentealba:2023rvf,Fuentealba:2023syb,Fuentealba:2025ekj,Gonzalez:2023yrz}, and often turn out to be related to `dual' symmetries put forward in earlier literature \cite{Strominger:2015bla,Hosseinzadeh:2018dkh,Godazgar:2018qpq,Godazgar:2019dkh,Kol:2019nkc,Porrati:2020anb,Javadinezhad:2022hhl}. The resulting Lie algebras contain infinite-dimensional versions of the Heisenberg algebra, whose irreducible unitary representations typically contain Fock spaces of soft gauge bosons. As we shall see, it is then a simple matter to identify the resulting representations with Hilbert spaces of dressed states.

The remainder of this work is organized in two parts: a brief review of the relevant extended asymptotic symmetries of electrodynamics and gravity, followed by a group-theoretic argument to prove both the insufficiency of asymptotic symmetries without extensions, and the success of suitable Heisenberg-extended symmetries. As should be clear from this introduction, the vast majority of the ingredients needed for the argument have already appeared in scattered portions of the literature. Our goal is merely to put those pieces together and explain, in a single stroke, how dressed states can be viewed as irreducible representations. This is ultimately a simple story, but one that we felt is worth being told.

\section{Asymptotic symmetries and their logarithmic duals}
\label{se2}

The asymptotic symmetries of $(3+1)$-dimensional electrodynamics and gravity have a similar structure, so we review their salient features here in one go. Crucially, we also include symmetry generators that are canonically conjugate to large gauge transformations and supertranslations (respectively) \cite{Fuentealba:2023syb}. The resulting Heisenberg extension will be of paramount importance for dressed states in section \ref{se3}. Note that field-theoretic details will often be unimportant for us, by contrast with the crucial group (or Lie algebra) structure. As a result, we will not need explicit expressions for surface charges, and refer instead to \cite{Fuentealba:2022xsz,Fuentealba:2023rvf,Fuentealba:2025ekj} for details.

\medskip\noindent\textbf{Electrodynamics.} We begin with electrodynamics, described by a U(1) gauge field $A_{\mu}$ on Minkowski spacetime. The field satisfies suitable logarithmic fall-off conditions at either spatial infinity \cite{Fuentealba:2022xsz,Fuentealba:2023rvf} or null infinity \cite{Fuentealba:2025ekj}, as well as certain parity conditions, all of which we omit. The resulting asymptotic symmetries are gauge transformations that do not fall off to the identity at large distances. Denoting by $x \in S^2$ a point on the sphere at infinity, asymptotic symmetries are typically specified by (smooth) functions of $x$. With logarithmic falloffs, asymptotic symmetries turn out to be generated by canonical charges of two kinds.

The first are standard asymptotic symmetries $Q_{\epsilon}$, each of which is labelled by an arbitrary real function $\epsilon(x)$ \cite{He:2014cra,Henneaux:2018gfi}. Abstractly, one can write any such charge as an integral over the sphere at infinity,
\be
Q_\epsilon
=
\oint_{S^2}\text{d}^2x\sqrt{g(x)}\,\epsilon(x)\,F(x),
\label{s10}
\ee
where $\text{d}^2x\sqrt{g(x)}$ is the usual measure on a unit sphere and $F(x)$ was defined in~\cite[eq.~(6.57)]{Fuentealba:2023rvf}: it is the sum of the radial gauge field and the retarded time derivative of the logarithmic Goldstone, generated by a pure logarithmic gauge transformation of $A$. The charge associated with constant $\epsilon$ is proportional to the electric charge of the system, while higher harmonics of $\epsilon$ are genuine \i{large gauge transformations}. The Poisson brackets of such charges vanish: $\{Q_{\epsilon},Q_{\epsilon'}\}=0$ for any two functions $\epsilon,\epsilon'$. They therefore span the Abelian Lie algebra $C^{\infty}(S^2)$ of functions on the sphere.

The second are logarithmic asymptotic symmetries $\widetilde Q_{\eta}$, labelled by an arbitrary function $\eta(x)$ with zero average (\ie without zero-mode in its harmonic expansion). These logarithmic charges also commute with one another, and they can be written as integrals
\be
\widetilde Q_{\eta}
=
\oint_{S^2}\text{d}^2x\sqrt{g(x)}\,\eta(x)\,\widetilde F(x),
\label{s11}
\ee
where $\widetilde F(x)$ was defined in~\cite[eq.~(6.59)]{Fuentealba:2025ekj} and is proportional to the Goldstone that is a pure large gauge transformation of $A$. The key insight of \cite{Fuentealba:2022xsz,Fuentealba:2023rvf} is that the standard charges \eqref{s10} and their logarithmic partners \eqref{s11} fail to commute: 
\be
\{Q_{\epsilon},\widetilde Q_{\eta}\}
=
\int_{S^2}\text{d}^2x\sqrt{g(x)}\,\epsilon(x)\,\eta(x).
\label{t11}
\ee
The latter therefore span an infinite-dimensional Heisenberg algebra, with generators of large gauge transformations and logarithmic transformations seen as functional analogues of canonically conjugate `position' and `momentum' operators. We shall denote this algebra as $\mathfrak{u}(1)\oplus\widehat{\text{LGT}\oplus\text{LGT}^*}$, where the acronym LGT refers to large gauge transformations and LGT$^*$ refers to their logarithmic, canonically conjugate partners. The hat serves to stress the presence of the Heisenberg central extension \eqref{t11}. The $\mathfrak{u}(1)$ factor is the overall electric charge, generated by the zeroth harmonic of $F(x)$ in \eqref{s10}, which commutes with everything by virtue of \eqref{t11}.

Finally, to complete the asymptotic symmetry algebra, one needs to know the commutators of the charges \eqref{s10}--\eqref{s11} with generators of the Poincaré group. While both sets of charges are translation-invariant, their relation to Lorentz transformations depends on the definition of the energy-momentum tensor of the theory. The non-symmetric energy-momentum tensor that follows from Noether's theorem yields Lorentz charges that act as conformal transformations of the sphere $S^2$ (see \eg \cite{Oblak:2015qia}), which induces nontrivial Lorentz transformation laws for the fields $\epsilon$, $\eta$, $F$ and $\widetilde F$. Namely, $\epsilon$ and $\widetilde F$ are densities with weight zero, while $F$ and $\eta$ are densities with weight $1$. The notion of `weight' is used here in the sense of Radon densities (see \eg \cite[sec.~2.2]{Bekaert:2022ipg}): a (real) function $H(x)$ on $S^2$ is a density with weight $w$ if it transforms as
\be
\big(\phi\cdot H\big)\big(\phi(x)\big)
\equiv
\bigg(\frac{g(x)}{g(\phi(x))}\bigg)^{w/2}
\bigg|\frac{\der\phi(x)}{\der x}\bigg|^{-w}
H(x)
\label{s25}
\ee
under any diffeomorphism $x\mapsto\phi(x)$ of $S^2$. The fact that both combinations $\epsilon(x)F(x)$ and $\eta(x)\widetilde F(x)$ have total weight 1 ensures that the charges \eqref{s10}--\eqref{s11} are diffeomorphism-invariant, \eg in the sense that $\oint\text{d}^2x\sqrt{g(x)}\epsilon(x)F(x)=\oint\text{d}^2x\sqrt{g(x)}(\phi\cdot\epsilon)(x)(\phi\cdot F)(x)$.

Instead of the non-symmetric energy-momentum tensor, one may choose to work with the symmetric one. The latter is gauge-invariant and therefore integrates to Lorentz charges that commute with all gauge charges. The two choices of Lorentz generators differ by a surface integral that is bilinear in $A_{\mu}$. The elimination of this surface integral in favour of the symmetric energy-momentum tensor's charge is commonly known as the (Belinfante-Rosenfeld) `improvement' procedure, see \eg\cite{Gieres_EMT}. Remarkably, it was observed in~\cite{Fuentealba:2023rvf} that this improvement is a canonical transformation when logarithmic gauge transformations are added to the symmetry algebra. For the sake of completeness and in order to compare with gravity, where there is no standard definition of an energy-momentum tensor, we will nonetheless consider both the semi-direct and direct product realizations of the asymptotic symmetry algebra of electromagnetism.

It is worth noting that Heisenberg central extensions of asymptotic symmetries have also been encountered in the discussion of `magnetic' (as opposed to `electric') charges \cite{Strominger:2015bla,Hosseinzadeh:2018dkh}. We stress that the structure \eqref{t11} holds regardless of any such interpretation. Anticipating the language of Hilbert spaces in section \ref{se3}, the only key point is that the operators $F(x)$ and $\widetilde F(x)$ both appear in the operator algebra of the theory, and that their commutator yields a Heisenberg extension \eqref{t11}. Whether $\widetilde F(x)$ is interpreted as a magnetic charge aspect, or merely as an operator canonically conjugate to $F(x)$ that stems from logarithmic falloffs, is ultimately irrelevant.

\medskip\noindent\textbf{Gravity.} The structure of logarithmic asymptotic symmetries in gravity is directly analogous to the electromagnetic setup, up to additional clutter with indices \cite{Fuentealba:2022xsz}. Thus, the spacetime metric $g_{\mu\nu}$ satisfies suitable fall-off conditions at spatial or null infinity, plus parity conditions, all of which we omit once again. The resulting asymptotic symmetries are again specified by functions or vector fields on a sphere $S^2$ at (spatial or null) infinity. With the logarithmic falloffs of \cite{Fuentealba:2022xsz}, asymptotic symmetries of gravity turn out to be generated by canonical charges of two kinds.

The first are either Lorentz charges or supertranslation charges. There are six linearly independent Lorentz charges, measuring the angular momentum and the centre-of-mass charges of the system. As for supertranslation charges, they are labelled by infinitely many parameters, namely the harmonics of an arbitrary real function $\epsilon(x)$ on the sphere at infinity \cite{Bondi,Sachs1,Sachs2,Henneaux:2018cst}. Each supertranslation charge is an integral over the sphere at infinity analogous to \eqref{s10}:
\be
Q_\epsilon
=
\oint_{S^2}\text{d}^2x\sqrt{g(x)}\,\epsilon(x)\,G(x),
\label{t13}
\ee
where $G(x)$ is defined in~\cite{ZwikelToAppear}: it is the sum of the radial gravitational field and of the logarithmic Goldstone, generated by a pure logarithmic supertranslation. The charges associated in this way with the four lowest harmonics of $\epsilon(x)$ correspond to the energy-momentum vector of the system, since the four lowest harmonics of $\epsilon$ (with angular momentum $\ell=0,1$) are interpreted as global spacetime translations. By contrast, the charges corresponding to higher harmonics of $\epsilon$ are genuine supertranslations, which do not occur in special relativity. Similarly to large gauge transformations in electrodynamics, supertranslations span an Abelian algebra of functions on $S^2$.

The second are logarithmic asymptotic symmetries $\widetilde Q_{\eta}$, labelled by an arbitrary function $\eta(x)$ whose four lowest harmonics may be taken to vanish (they do not contribute to any charge) \cite{Fuentealba:2022xsz}. Again, such logarithmic charges commute with one another, and they can be written as
\be
\widetilde Q_{\eta}
=
\oint\text{d}^2x\sqrt{g(x)}\,\eta(x)\,\widetilde G(x)
\label{s14}
\ee
where $\widetilde G(x)$ is defined in~\cite{ZwikelToAppear} and is proportional to the Goldstone that is a pure supertranslation of $g_{\mu\nu}$. The key insight of \cite{Fuentealba:2022xsz} is that the (super)translation charges \eqref{t13} and their logarithmic partners \eqref{s14} fail to commute: their bracket involves a Heisenberg central extension of the same form as in eq.~\eqref{t11}. The set of sums of supertranslation charges \eqref{t13} and their logarithmic duals \eqref{s14} thus spans an infinite-dimensional Heisenberg algebra. We denote this algebra as $\RR^4\oplus\widehat{\text{ST}\oplus\text{ST}^*}$, where $\text{ST}$ and $\text{ST}^*$ respectively refer to supertranslations (with vanishing harmonics $\ell=0,1$) and their logarithmic duals, while the hat stresses again the presence of the Heisenberg central extension \eqref{t11}. The $\RR^4$ factor is the commuting subalgebra of spacetime translations, which are the gravitational analogues of the global U(1) charge in electrodynamics.

It remains to determine the commutation relations of the charges \eqref{t13}--\eqref{s14} with the Lorentz generators. By the same analysis as above, one finds that the latter act as conformal transformations of $G(x)$ and $\tilde{G}(x)$ on the celestial sphere. More specifically, (super)translations turn out to be densities with weight $(-1/2)$, so that the field $G(x)$ in eq.~\eqref{t13} is a density with weight $3/2$ on $S^2$, in the sense of eq.~\eqref{s25}. This is a functional generalization of the vector representation of the Lorentz group. Conversely, the canonically conjugate field $\widetilde G(x)$ of eq.~\eqref{s14} transforms as a density with weight $(-1/2)$,  ensuring again that the pairing between $G$ and $\widetilde G$ is Lorentz-invariant. The logarithmic supertranslation $\eta(x)$ has weight $3/2$. Finally, note again the observation of~\cite{Fuentealba:2023syb}, that supertranslations can be decoupled from the Lorentz generators by a canonical transformation of the symmetry algebra in the presence of logarithmic supertranslations. We show below that an analogous decoupling occurs in the Hilbert space given by representation theory, without requiring any redefinition of generators.

\section{Dressed states as irreducible representations}
\label{se3}

As outlined in the introduction, our goal is to show that logarithmic asymptotic symmetries can be used to build symmetry groups whose irreducible unitary representations provide Hilbert spaces of dressed states. We now describe this for both electrodynamics and gravity, and point out along the way why asymptotic symmetries \i{without} their canonical duals do not suffice.

Consider \eg a massive, electrically charged particle. As follows from Wigner's seminal classification \cite{Wigner:1939cj}, one may view the Hilbert space of this `naked' particle as the carrier space of an irreducible, unitary representation of the (connected) Poincar\'e group
\be
P
=
\text{SO}(3,1)^{\uparrow}\ltimes\RR^4,
\label{s15}
\ee
with some mass and some spin. The quantum states of the particle are wave functions in space or momentum space; those of a massive particle with spin $s$ belong to the Hilbert space $L^2(\RR^3)\otimes\CC^{2s+1}$. Standard scattering states with definite momentum are essentially plane waves in that Hilbert space.

The key insight of \cite{Kulish,Chung:1965zza,Kibble:1968sfb,Kibble:1968oug,Kibble:1968npb,Kibble:1968lka} is that such `naked' scattering states give rise to infrared-divergent scattering amplitudes due to the long-range interactions carried by gauge bosons, such as those of electrodynamics or gravity. As it turns out, such divergences are cured by accepting that asymptotic states are never genuinely `naked', and always carry along a cloud of soft gauge bosons \cite{Kulish}. This `infrared dressing' is ultimately rather natural in quantum field theory, but it is revolutionary from a representation-theoretic perspective: it means that the Hilbert space of a dressed particle is much, much larger than the measly $L^2(\RR^3)$ of a naked particle, since dressed particles can probe a full Fock space of soft gauge bosons. The question is: what symmetry group, if any, has irreducible unitary representations that contain such Fock spaces?

\medskip\noindent\textbf{Poincar\'e-Maxwell and BMS.} The usual Poincar\'e group \eqref{s15} is manifestly insufficient to incorporate soft degrees of freedom. So is its trivial extension $P\times\text{U}(1)$, where the $\text{U}(1)$ factor comprises global, `rigid' gauge transformations that commute with all Poincar\'e transformations. Now, consider what happens when the Poincar\'e group \eqref{s15} is extended not by U(1), but by the full group of large gauge transformations \eqref{s10}. The latter are functions on a sphere, acted upon by Lorentz transformations seen as conformal transformations of $S^2$ through the usual isomorphism $\text{SO}(3,1)^{\uparrow}\cong\text{SL}(2,\CC)/\ZZ_2$ (see \eg~\cite{Oblak:2015qia}). The extended symmetry thus becomes the `Poincar\'e-Maxwell group'
\be
P\text{Max}
\equiv
\text{SO}(3,1)^{\uparrow}\ltimes\Big(\RR^4\times\text{U(1)}\times\text{LGT}\Big),
\label{s17}
\ee
which is now infinite-dimensional and involves a genuine nontrivial action of Lorentz transformations on large gauge transformations. [As in section \ref{se2}, we separate the global U(1) factor from genuine large gauge transformations.] This is a lot more promising if one is hoping to obtain dressed states! Indeed, \eqref{s17} is essentially an analogue of the global BMS group in gravity, which similarly takes the form \cite{Bondi,Sachs1,Sachs2}
\be
\text{BMS}
=
\text{SO}(3,1)^{\uparrow}\ltimes\Big(\RR^4\times\text{ST}\Big),
\label{t17}
\ee
where we distinguish once again between spacetime translations and genuine supertranslations. Such groups have actually been related to dressed states in the literature, because the action of Lorentz transformations on null infinity induces their nontrivial action on dressing operators \cite{Choi:2017ylo}.

The problem with \eqref{s17} and \eqref{t17} is that they are not large enough to fully accommodate the soft Fock space needed for dressed states. For example, consider the Poincar\'e-Maxwell group \eqref{s17} and ask how its irreducible unitary representations look like. The group is a semidirect product with an Abelian normal subgroup, so its unitary representations are guaranteed \cite{mackey1968induced} to take the same form as those of the Poincar\'e group \eqref{s15} \cite{Wigner:1939cj}, or the BMS group \eqref{t17} \cite{McCarthy00,McCarthy01,McCarthy317,McCarthy301}. Namely, one starts by choosing a `momentum vector' in the dual space of the Abelian group, then one finds the orbit $\cO$ of that momentum under Lorentz transformations. This orbit is essentially the manifold on which the wave functions of the carrier space $L^2(\cO)$ live. In the simplest (scalar) case, this is the end of the story; for particles with spin, the construction is slightly more involved (wave functions are vector-valued), but the fact remains that wave functions live on an orbit of `momenta' under Lorentz transformations.

So, how would such representations be built for the group \eqref{s17}? A `momentum' in that case is a pair $(p,F)$, where $p=(p_0,p_1,p_2,p_3)$ is an actual energy-momentum vector, while $F=F(x)$ is an electric charge aspect giving rise to charges such as \eqref{s10}. The orbit on which a wave function lives is the set
\be
\cO_{(p,F)}
=
\Big\{
\big(\Lambda\cdot p,\Lambda\cdot F\big)\Big|\Lambda\in\text{O(3,1)}
\Big\}
\label{s18}
\ee
of all pairs obtained by acting on $(p,F)$ with Lorentz transformations. [Specifically, the transformation $p\mapsto\Lambda\cdot p$ is given by usual matrix multiplication, while the transformation $F\mapsto\Lambda\cdot F$ is given by the action \eqref{s25} of conformal transformations on a density with weight $w=1$ when $\Lambda$ is identified with the diffeomorphism it induces on $S^2$.] Unsurprisingly, such an orbit is more complicated than the standard `mass shell' of a relativistic momentum vector, and the number of Lorentz-inequivalent orbits of this kind is enormous (the moduli space of orbits is infinite-dimensional). But an elementary property of \i{any} orbit of the form \eqref{s18} is that it is at most six-dimensional, since it is a homogeneous space under the six-dimensional Lorentz group. As a result, the carrier space $L^2(\cO)$ of any representation built in this way consists of wave functions on a finite-dimensional manifold; there is no room there for soft gauge bosons.

The same conclusion holds, in gravity, for the global BMS group \eqref{t17}, whose irreducible unitary representations are indeed known to consist of wave functions on finite-dimensional orbits \cite{McCarthy00,McCarthy01,McCarthy317,McCarthy301}. The orbits in that case are different from \eqref{s18} because Bondi mass aspects (dual to supertranslations) have weight $3/2$ rather than $1$ under conformal transformations of $S^2$, but the fact remains that the resulting Hilbert space $L^2(\cO)$ leaves no room for soft gravitons.

\medskip\noindent\textbf{Extended Poincar\'e-Maxwell and BMS.} The above is the sense in which spacetime isometries together with `standard' asymptotic symmetries fall short of providing Hilbert spaces of dressed quantum states. The way out is the one announced in sections \ref{se1} and \ref{se2}: further extensions of the symmetry group---and in particular central extensions---are needed in order to go beyond the one-particle Hilbert space $L^2(\cO)$ obtained from the Wigner construction. Specifically, consider the extensions of \eqref{s17} and \eqref{t17} given by
\begin{align}
\label{s20}
\!\!\widehat{P\text{Max}}
&\equiv
\text{SO}(3,1)^{\uparrow}\!\ltimes\!\Big(\RR^4\!\times\!\text{U}(1)\!\times\!\widehat{\text{LGT}\!\times\!\text{LGT}^*}\Big)\\
\label{t20}
\!\!\widehat{\text{BMS}}
&\equiv
\text{SO}(3,1)^{\uparrow}\!\ltimes\!\Big(\RR^4\times\widehat{\text{ST}\!\times\!\text{ST}^*}\Big)
\end{align}
where the hatted factors on the right-hand side are the Heisenberg groups of section \ref{se2} whose algebras take the form \eqref{t11}. An element of $\widehat{P\text{Max}}$ is thus a  5-tuple $(\Lambda,a,\epsilon,\eta,\lambda)$ consisting of a Lorentz transformation $\Lambda$, a spacetime translation $a$, a large gauge transformation $\epsilon(x)$ that generally has a nonvanishing zero-mode, a logarithmic large gauge transformation $\eta(x)$ without zero-mode, and a real number $\lambda$ playing the role of a dual to the Heisenberg central charge. Similarly, an element of $\widehat{\text{BMS}}$ is a 4-tuple $(\Lambda,\epsilon,\eta,\lambda)$ where $\epsilon(x)$ is a supertranslation that generally has all harmonics nonzero, while $\eta$ is a dual supertranslation where the only nonzero harmonics are those with $\ell\geq2$.

Having introduced the groups \eqref{s20}--\eqref{t20}, let us ask again how their irreducible unitary representations look like. These groups are semi-direct products, but their normal subgroup is \i{not} Abelian, so Mackey's theorem on induced representations does not apply \cite{mackey1968induced}. In fact, we are not aware of a general theorem that provides the classification of all irreducible unitary representations of groups of this form. It is nevertheless straightforward to apply the same logic of induced representations. Namely, focus for definiteness on the electromagnetic group \eqref{s20} and let $\cH$ be the carrier space of an irreducible, unitary representation of that group. It is certainly the case that the subgroup $\RR^4\times\text{U}(1)\times\widehat{\text{LGT}\!\times\!\text{LGT}^*}$ acts unitarily on $\cH$; at the very least, this action is given by one of its irreducible unitary representations. Any such representation is a tensor product of a one-dimensional representation of $\RR^4\times\text{U}(1)$, labelled by some momentum $p$ and some charge $q$, with an irreducible unitary representations of  $\widehat{\text{LGT}\!\times\!\text{LGT}^*}$. The latter is an infinite-dimensional Heisenberg group. If it were finite-dimensional, the Stone-von Neumann theorem would guarantee the uniqueness of its irreducible unitary representation, given any nonzero value for its central charge. The theorem fails to hold in infinite dimension---this is, after all, one of the main subtleties in quantizing field theories. 

 We expect the physically relevant representation to be the Hilbert space of a field theory on the celestial sphere that describes the zero-frequency, soft photons. The same conclusion involving soft gravitons stems from the analysis of the extended BMS group \eqref{t20}. In all explicit constructions of dressed states in the literature, these soft Hilbert spaces arise from the Fock space of a Gaussian field, where LGT eigenstates are (generalized) coherent states obtained by the action of $\text{LGT}^*$ on the Fock space vacuum. The correct formulation of a Hilbert space that contains these coherent states is an open problem (see \eg\cite{Nonsep_Prabhu}) that we will not address here. Rather, we will view the soft Hilbert space abstractly as $L^2(\text{LGT}^*)$, \ie a space of square-integrable functionals of the electric charge aspect $F(x)$.

Thus, the structure of the groups \eqref{s20}--\eqref{t20} ensures that their irreducible unitary representations contain a full Hilbert of soft gauge bosons. There is more: the inclusion of logarithmic symmetries in \eqref{s20}--\eqref{t20} drastically simplifies the classification of irreducible representations, despite making the groups larger. Indeed, consider once more the carrier space $\cH$ of some irreducible unitary representation of the electromagnetic group \eqref{s20}, let $p$ be a momentum vector and $q$ a U(1) charge, and assume that the pair $(p,q)$ appears in the decomposition of $\cH$ in irreducible representations of the subgroup $\RR^4\times\text{U}(1)\times\widehat{\text{LGT}\!\times\!\text{LGT}^*}$. What other momenta (and charges) appear in $\cH$, given that the Lorentz group acts nontrivially on the subgroup $\RR^4\times\text{U}(1)\times\widehat{\text{LGT}\!\times\!\text{LGT}^*}$? As far as momenta alone are concerned, the answer is the same as for a standard naked particle: the set $\cO$ of momenta reached by acting on $p$ with Lorentz transformations is a `mass shell'. [The global U(1) charge $q$ is Lorentz-invariant, since Lorentz transformations act trivially on the U(1) factor in \eqref{s20}.] But this is now the end of the story: Lorentz transformations do act on large gauge transformations and their duals, hence on soft photon quantum states in a representation of $\widehat{\text{LGT}\!\times\!\text{LGT}^*}$. However, since the action of the Lorentz group preserves the carrier space as a whole, its `orbit' consists of a single point. As a result, the only label of the irreducible representation of $\widehat{\text{LGT}\!\times\!\text{LGT}^*}$ is the central charge, which is Lorentz-invariant.

The end result is that the irreducible unitary representations of, say, the extended BMS group \eqref{t20}, are labelled exactly by the same quantities as irreducible representations of Poincar\'e---namely mass and spin. The irreducible representations of the extended Poincar\'e-Maxwell group \eqref{s20} are similarly labelled by their mass and spin, plus the global electric charge. In this sense, adding dual asymptotic symmetries as in \eqref{s20}--\eqref{t20} simplifies the classification of representations of the underlying groups, since orbits of mass aspects or charge aspects such as \eqref{s18} are no longer needed. The `price' to pay---really a perk in fact---is that the Hilbert space of such a representation is much larger than that of an irreducible representation of the Poincar\'e group. For example, the Hilbert space of a naked scalar massive particle is $L^2(\RR^3)$, while the corresponding carrier space of an irreducible unitary representation of \eqref{t20} is $\cH=L^2(\RR^3)\otimes L^2(\text{ST}^*)$, where $L^2(\text{ST}^*)$ is the Fock space of soft gravitons---again with the aforementioned difficulty linked to the definition of a measure on $\text{ST}^*$. A vector in $\cH$ is thus an admissible quantum state for a (massive, scalar) particle dressed with soft gravitons.

The simplification of representation theory is reminiscent of a similar algebraic simplification when logarithmic transformations are included. Indeed, it was shown in \cite{Fuentealba:2023syb,Fuentealba:2022xsz,Fuentealba:2023rvf,Fuentealba:2025ekj} that the Lie algebra of the groups \eqref{s20}--\eqref{t20} is such that a nonlinear redefinition of generators makes all Poincar\'e transformations commute with genuine (high-harmonic) asymptotic symmetries. This factorized structure in the universal enveloping algebra ultimately entails factorization at the level of a Hilbert space such as $\cH=L^2(\RR^3)\otimes L^2(\text{ST}^*)$. It is also responsible for the simplification of the classification of representations, since representations of the extended groups \eqref{s20}--\eqref{t20} are ultimately classified by the same labels as those of representations of the Poincar\'e group.

\medskip
\medskip
\section{Conclusion}

We have thus found symmetry groups \eqref{s20}--\eqref{t20} whose irreducible unitary representations are labelled by the usual mass and spin (and possibly electric charge), and whose carrier spaces are spanned by the quantum states of a relativistic particle dressed with soft gauge bosons. This is our main observation in this work.

In principle, further investigation is needed to determine how relevant the groups \eqref{s20}--\eqref{t20} are for scattering and/or the quantization of gauge theories. It may be worth stressing that, while dressed state constructions formally solve the infrared divergence problem, they are almost never used in practice. In particular, practical computations are based on inclusive cross-sections. The fact that dressings require cutoffs to be well-defined also obscures the analytic properties of the corresponding S-matrix elements, which is a subject of active research (see \eg \cite{Bellazzini:2025bay}). A better understanding of their group-theoretic underpinnings may therefore help make progress on these fronts.

Finally, another natural next step is to study the coadjoint representation and orbits of the groups \eqref{s20}--\eqref{t20} with the same motivations as in earlier literature \cite{Barnich2,Barnich:2021dta}. In the case of the extended BMS group \eqref{t20}, this leads to the inhomogeneous transformation law of the asymptotic shear under supertranslations, which is effectively a restatement of the central extension \eqref{t11}. One can then push the computation further and find \eg the corresponding geometric actions, which are expected to describe perturbative boundary dynamics \cite{BarnichGeom,OblakThesis,Cotler,Barnich:2022bni}.

\section*{Acknowledgements}

This project began many years ago and evolved adiabatically for a long time, during which we benefited from fruitful discussions with a number of colleagues. We are grateful to Glenn Barnich, Mathieu Beauvillain, Xavier Bekaert, Sangmin Choi, Adrien Fiorucci, Marc Geiller, Mahdi Godazgar, Hern\'an Gonz\'alez, Marc Henneaux, Olivera Mi\v{s}kovi\'c, Kevin Nguyen, Romain Ruzziconi, Jakob Salzer, Michele Schiavina, Ali Seraj and C\'eline Zwikel for many such interactions, and for their support in this long endeavour. The work of J.B. was partially supported by the Swiss National Science Foundation through the NCCR SwissMAP. JAM acknowledges funding from the
European Union (ERC “QFTinAdS”, project number 101087025).

\providecommand{\href}[2]{#2}
\end{document}